\documentclass[12pt]{iopart}
\usepackage{graphicx}
\usepackage{iopams}
\usepackage{subfigure}
\usepackage{color}
\usepackage{hyperref}
\usepackage{stmaryrd}
\newcommand{\mbr}{\mathbf{r}}
\newcommand{\beq}{\begin{equation}}
\newcommand{\eeq}{\end{equation}}

\begin{document}

\title{Quantum scaling of the spin lattice relaxation rate in the checkerboard $J$-$Q$ model}

\author{Chengchen Li$^{1}$, Huihang Lin$^{1}$, and Rong Yu$^{1,2,\dagger}$}
\address{$^{1}$ Department of Physics and Beijing Key Laboratory of Opto-electronic Functional Materials \& Micro-nano Devices, Renmin University of China, Beijing 100872, China}

\address{$^{2}$ Key Laboratory of Quantum State Construction and Manipulation (Ministry of Education), Renmin University of China, Beijing, 100872, China}
\address{$^{\dagger}$ rong.yu@ruc.edu.cn}

\begin{abstract}
Motivated by recent progress on the experimental realization of proximate deconfined quantum critical point in a frustrated quantum magnet, we study the low-energy spin dynamics of a related checkerboard $J$-$Q$ model by using quantum Monte Carlo simulations. The ground state of this model undergoes a weakly first-order quantum phase transition with an emergent $O(4)$ symmetry between an antiferromagnetic state and a plaquette valence bond solid. The calculated spin lattice relaxation rate of nuclear magnetic resonance, $1/T_1$, exhibits distinct low-temperature behaviors depending on the ground states. With decreasing the temperature, $1/T_1$ rises up on the antiferromagnetic side, characterizing a crossover to the renormalized classical regime, whereas $1/T_1$ drops exponentially on the side of valence bond solid, reflecting the gap opening in the plaquette ordered phase. The extracted spin gap scales with the distance to the transition point as a power-law with an exponent $\phi\approx0.3$, consistent with the scaling ansatz $\phi=\nu z$ with $\nu\approx0.3$ and $z=1$. Near the quantum phase transition, the temperature dependent $1/T_1$ shows a broad crossover regime where a universal scaling $1/T_1\sim T^{\eta}$ with $\eta\approx0.6$ is found. Our results suggest a quantum scaling regime associated with the emergent enhanced symmetry near this first-order quantum phase transition.
\end{abstract}


\submitto{\JPCM}
\maketitle

\section{Introduction}

Order-disorder physics is one major theme in quantum magnetism and has been extensively studied in spin frustrated systems, where a wealth of intriguing phenomena has been discovered~\cite{Diep_book_2004, Mila_book_2010}. The quantum spin liquid (QSL) is an exotic, highly entangled quantum disordered state that supports deconfined fractionalized spin excitations~\cite{Balents_Nature_2010, Savary_RPP_2017, Zhou_RMP_2017}. Although QSLs have been shown to be the exact ground states of certain theoretical models, it is still under debate on whether they can be realized in frustrated magnetic materials.

Besides QSLs, theory predicted that deconfined fractional spin excitations can also arise at the so called deconfined quantum critical point (DQCP) that separates two magnetically ordered states with different symmetries~\cite{Senthil_science_2004, Senthil_PRB_2004}. As an example, the quantum phase transition (QPT) between an antiferromagnetic (AFM) state and a valence bond solid (VBS) is predicted to be continuous within this theory, and the transition point is called a DQCP. At this point, deconfined fractional spin excitations emerge, and the AFM and VBS order parameters are bound to a super vector with emergent enhanced symmetry. This scenario is in direct contrast to the QPT between two unrelated ordered states, for which the transition would be first-order according to the Landau theory of phase transitions. Lots of theoretical efforts have been made for studying this beyond-Landau QPT~\cite{Nahum_PRX_2015, Nahum_PRL_2015, Shao_science_2016, Wang_PRX_2017, Ma_PRB_2018, Lee_PRX_2019, Zhao_PRL_2020,Sreejith_PRL_2019, Serna_PRB_2019,   Takahashi_PRR_2020, Zhang_PRB_2020,Lu_PRB_2021,Liu_SB_2022, Xi_PRB_2023,  Yang_PRB_2022, Liu_PRX_2022, Shackleton_PRB_2021}, and the $J$-$Q$ model and its variants are one class of lattice models that have been extensively investigated~\cite{Shao_science_2016, Ma_PRB_2018, Sandvik_PRL_2007, Zhao_NP_2019}. The Hamiltonian of the $J$-$Q$ model contains nearest neighboring AFM Heisenberg interactions and ring-exchange-like four-spin interactions around plaquettes of the square lattice~\cite{Sandvik_PRL_2007}. This model and its variants generically holds a QPT between an AFM state and a plaquette valence bond solid (PVBS). There are numerical evidences that DQCPs with emergent $SO(5)$ and $O(4)$ symmetries can be realized in the $J$-$Q$ model and its easy-plane variant, respectively~\cite{Ma_PRB_2018,Sandvik_PRL_2007, Qin_PRX_2017, Sandvik_CPL_2020}.

Experimental evidence for DQCP has been recently reported by a nuclear magnetic resonance (NMR) study on a quantum magnet, SrCu$_2$(BO$_3$)$_2$.\cite{Cui_Science_2023} This compound can be well described by a spin frustrated Shastry-Sutherland model~\cite{Kageyama_PRL_1999,Miyahara_PRL_1999,Miyahara_JPCM_2003,Shastry_PBC_1981, Koga_PRL_2000, Corboz_PRB_2013,Kodama_JPCM_2005,Kageyama_JPSJ_1998}, and a PVBS ground state is stabilized by applying a pressure above about 2GPa~\cite{Zayed_NP_2017, Guo_PRL_2020, Jimenez_Nature_2021}. It has been shown that further applying a magnetic field can drive a PVBS-AFM transition~\cite{Cui_Science_2023,Haravifard_NC_2016, Shi_NC_2022}. This transition is weakly first-order and quantum scaling of the NMR spin lattice relaxation rate $1/T_1$ emerging near the transition point provides evidence for proximity to a DQCP~\cite{Cui_Science_2023}.

Motivated by the experimental progress, we study the spin dynamics near the PVBS-AFM transition theoretically. Though the Shastry-Sutherland model can well describe the low-temperature physics of SrCu$_2$(BO$_3$)$_2$, the strong spin frustration of this model makes theoretical study on the nature of the transition difficult, and existing results are even controversial~\cite{Lee_PRX_2019, Xi_PRB_2023, Yang_PRB_2022, Corboz_PRB_2013, Keles_PRB_2022, Boos_PRB_2019, Wessel_PRB_2018}. We then consider an alternative checkerboard $J$-$Q$ (CBJQ) model~\cite{Zhao_NP_2019}, which keeps the essential symmetry of the Shastry-Sutherland lattice and also holds a AFM-PVBS transition. The advantage of the CBJQ model is that it is sign problem free, and can be studied via quantum Monte Carlo (QMC) simulations. The ground state phase diagram of the CBJQ model has been investigated and it is shown that the ground state evolves from an AFM state to a PVBS with increasing $Q$ via a first-order QPT~\cite{Zhao_NP_2019}. It is remarkable that an enhanced $O(4)$ symmetry emerges at this first-order transition. The thermodynamic properties of the CBJQ model have also been discussed in both 2D and 3D~\cite{Guo_PRL_2020, Sun_CPB_2021}.

In this paper, we study the low-energy spin dynamics of the CBJQ model near the AFM-to-PVBS transition by calculating the NMR spin lattice relaxation rate $1/T_1$. We show that the low-temperature behavior of $1/T_1$ reflects the nature of the corresponding ground state. With decreasing the temperature, $1/T_1$ increases when the ground state is AFM, whereas $1/T_1$ drops exponentially when the ground state is a gapped PVBS. The extracted spin gap scales with the distance to the transition point $Q-Q_c$ as a power-law, with an exponent $\phi\approx0.3$. Near the QPT, $1/T_1$ shows first a maximum then a minimum by decreasing the temperature, giving rise to a broad crossover regime in which $1/T_1\sim T^{\eta}$ with a universal exponent $\eta\approx0.6$. These results suggest a quantum scaling regime associated with the emergent enhanced symmetry near this QPT, despite its firs-order nature. The rest of the paper is organized as follows. The CBJQ model and the theoretical method used in this work are introduced in Sec.~\ref{Sec:Model}. We then present results of $1/T_1$ and the finite-temperature phase diagram based on these results in Sec.~\ref{Sec:PhD}. Next, we discuss the scaling properties of $1/T_1$ in Sec.~\ref{Sec:Scaling}. In Sec.~\ref{Sec:Conclusion}, we provide further discussions and draw conclusion.

\section{Model and method}\label{Sec:Model}

The CBJQ model is defined on a square lattice and the Hamiltonian reads
\begin{equation}
H=J\sum_{\langle i,j\rangle}\mathbf{S}_{i}\cdot\mathbf{S}_{j}
-Q\sum_{ijkl\in\square} \left(\mathbf{S}_{i}\cdot\mathbf{S}_{j}\right) \left(\mathbf{S}_{k}\cdot\mathbf{S}_{l}\right) +\left(\mathbf{S}_{i}\cdot\mathbf{S}_{k}\right)\left(\mathbf{S}_{j}\cdot\mathbf{S}_{l}\right),\label{eq:Ham}
\end{equation}
where $J>0$ is an AFM exchange coupling between nearest neighboring spins. The $Q$ term refers to a ring-exchange-like interaction for four spins involved in a plaquette, as illustrated in Fig.~\ref{fig:1}{\bf a}. In this work, we calculate the specific heat and the NMR spin lattice relaxation rate to study the finite-temperature properties of this model by using the stochastic series expansion (SSE) QMC simulation~\cite{Dorneich_PRE_2001, Sandvik_AIPCP_2010,Syljuasen_PRE_2002}. In the SSE algorithm, the specific heat is evaluated by
\begin{equation}
 C = \frac{d\langle H \rangle}{dT} = \frac{1}{N} \left[ \langle n^2 \rangle - \langle n \rangle^2 - \langle n \rangle \right],
\end{equation}
where $n$ refers to the SSE expansion order, and $\langle \cdots \rangle$ denotes ensemble average.

The NMR spin relaxation rate is defined as
\begin{equation}
 \label{Eq:T1Def}
	1/T_1^{\alpha \alpha}
	\sim \lim\limits_{\omega \rightarrow 0} T\sum_q \chi^{\prime \prime}_{\alpha \alpha}(\boldsymbol{q}, \omega)/\omega,
	\end{equation}
where $\chi^{\prime \prime}_{\alpha \alpha}(\boldsymbol{q}, \omega)$ is the imaginary part of the generalized dynamical susceptibility at momentum $\boldsymbol{q}$ with spin component $\alpha=x,y,z$. Instead of directly evaluating the dynamical susceptibility via complicated numerical analytical continuation, here we adopt an alternative approach by evaluating $1/T_1$ from the following approximation:
\begin{equation}\label{Eq:T1Approx}
1/T_1^{\alpha \alpha} \simeq \beta \sum_{\boldsymbol{q}} \langle S^\alpha_{-\boldsymbol{q}} (\tau=\beta/2) S^\alpha_{\boldsymbol{q}} (0) \rangle,
\end{equation}
where $S^\alpha_{\boldsymbol{q}} (\tau)$ denotes the $\alpha$-component of the spin operator at momentum $\boldsymbol{q}$ and imaginary time $\tau$, and $\beta=1/T$ is the inverse temperature. This approach has been proven to provide an efficient estimate of the spin lattice relaxation rate in a number of models~\cite{Randeria_PRL_1992,Fan_PRR_2020}. In SSE QMC, the imaginary correlator can be computed via the directed loop algorithm~\cite{Dorneich_PRE_2001, Sandvik_AIPCP_2010, Syljuasen_PRE_2002}. In this work, we calculate $1/T_1^{zz}$ only and denote $1/T_1=1/T_1^{zz}$ in the rest of the paper because the $SU(2)$ spin rotational symmetry is preserved in this model at finite temperatures.

{To accurately determine the thermal transition to the PVBS, we also calculate the Binder cumulant $U_p$ associated with the PVBS order parameter $m_p$,
\begin{equation}
 U_p = \frac{3}{2} \left( 1-\frac{\langle m_p^4 \rangle}{3\langle m_p^2 \rangle^2} \right),
\end{equation}
where 
\begin{equation}
 m_p = \frac{2}{L^2} \sum_{\mbr} \theta(\mbr) S^z_{\mbr} S^z_{\mbr+\hat{x}} S^z_{\mbr+\hat{y}} S^z_{\mbr+\hat{x}+\hat{y}},
\end{equation}
$\mbr$ runs over the lower-left corners of the $Q$ plaquettes, and $\theta(\mbr)=\pm 1$ correspond to even or odd plaquette rows.}

We run the QMC simulation on $N=L\times L$ square lattices with periodic boundary condition, and take $J=1$ as the energy unit. The largest linear system size in our simulations is $L=96$ and the lowest temperature accessed is $T/J=0.1$ {A typical simulation run includes $10^7$ MC steps, and errorbars are estimated by using standard method of bin statistics~\cite{Sandvik_AIPCP_2010}}.

\section{Finite-temperature results and the phase diagram}\label{Sec:PhD}

\begin{figure}[th!]
\centering\includegraphics[
width=1.0\textwidth
]{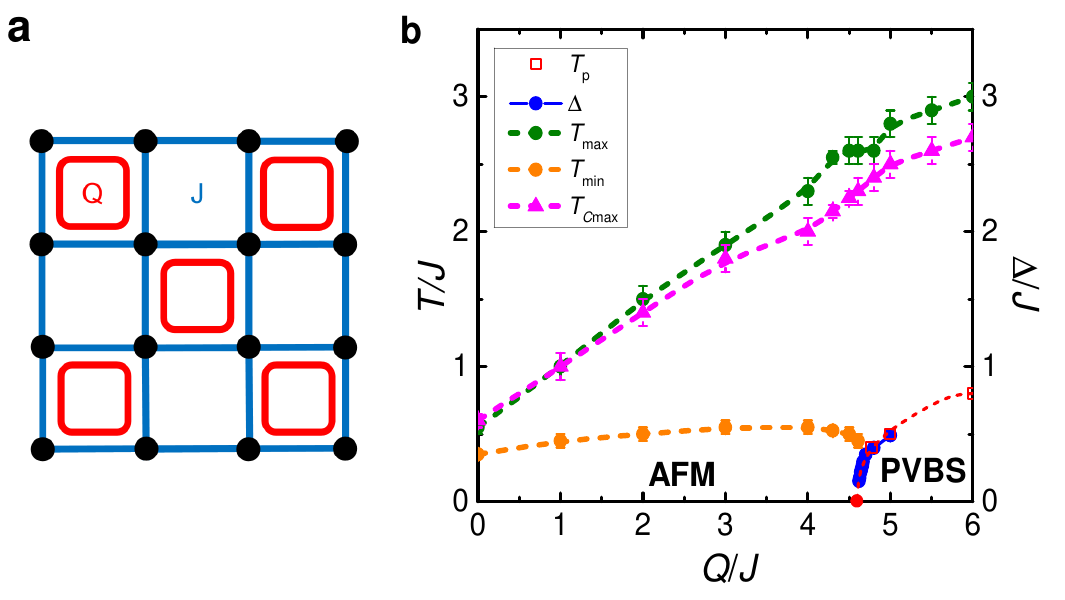}
\caption{(Color online) {\bf a} Sketch of the CBJQ model, where $S=1/2$ spins are located at sites of the square lattice, with nearest neighboring Heisenberg coupling $J$ and ring-exchange-like interaction $Q$ for spins around a circle. {\bf b} The finite-temperature phase diagram of the CBJQ model. At zero temperature, the antiferromagnetic (AFM) and plaquette valence bond solid (PVBS) ground states are separated by a first-order QPT with an emergent $O(4)$ symmetry (the red circle). The PVBS order survives finite temperature up to $T_{\rm p}$ (red squares), where the red dashed line is a guide to the eyes. The magenta dashed line ($T_{C\rm{max}}$) with triangles is a crossover determined from the maximum of the specific heat curve in Fig.~\ref{fig:2}{\bf a}. Orange and green dashed lines with circles are crossovers determined from minimum (at $T_{\rm{min}}$) and maximum (at $T_{\rm{max}}$) of the $1/T_1$ curve in Fig.~\ref{fig:2}{\bf b}, respectively. Blue circles are spin gaps extracted from the $1/T_1$ curve on the PVBS side.
}
\label{fig:1}
\end{figure}

\begin{figure}[th!]
\centering\includegraphics[
width=1.0\textwidth
]{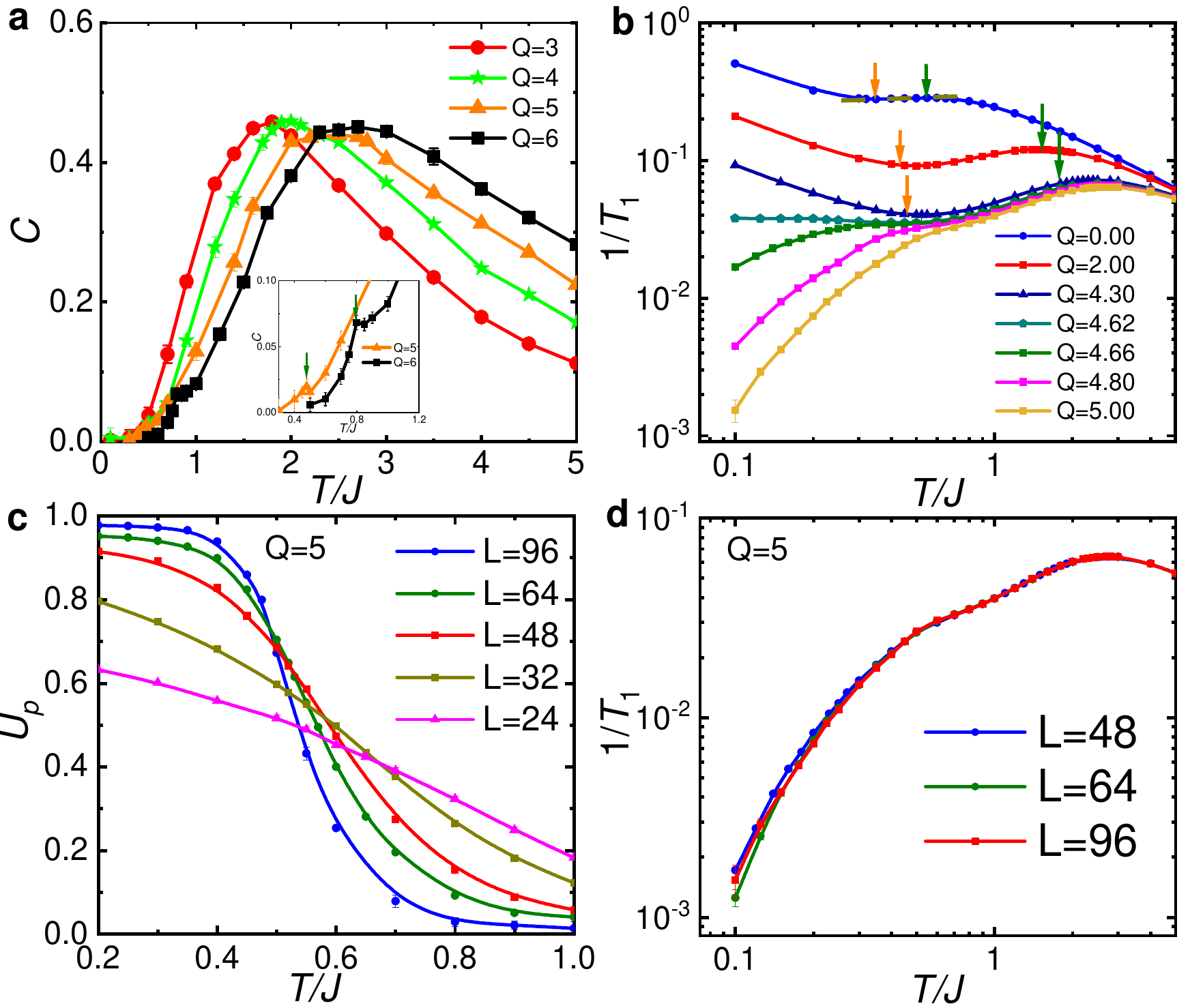}
\caption{(Color online) {\bf a} Temperature dependent specific heat $C$ at several $Q$ values. {The inset shows details of the $C(T)$ curve in the temperature window near the thermal transition to the PVBS phase for $Q>Q_c$. The arrow indicates the transition temperature $T_{\rm{p}}$.} {\bf b} Temperature dependent $1/T_1$ at several $Q$ values. The orange and green arrows indicate positions of maximum and minimum ($T_{\rm{min}}$ and $T_{\rm{max}}$) of each curve, respectively. The dark yellow line on top of the $Q=0$ data is a power-law fit of $1/T_1\sim T^\eta$ with the exponent $\eta=0.03$. {{\bf c} Finite-size scaling of the Binder cumulant $U_p$ associated with the PVBS order parameter $m_p$ for $Q=5$, where the transition temperature $T_{\rm{p}}$ can be determined from the crossing point of $U_p$. {\bf d} Temperature dependent $1/T_1$ at $Q=5$ for several finite-size systems.}
}
\label{fig:2}
\end{figure}

The ground state phase diagram of the CBJQ model has been investigated and it is shown that the ground state evolves from an AFM state to a PVBS with increasing $Q$ via a first-order QPT with an emergent enhanced $O(4)$ symmetry at $Q_c/J\approx4.6$.\cite{Zhao_NP_2019} Here we investigate the thermodynamics and spin dynamics at finite temperature of this model. Our main results are summarized in the phase diagram in Fig.~\ref{fig:1}{\bf b}. This phase diagram is based on the calculated temperature dependent specific heat $C$ and NMR spin lattice relaxation rate $1/T_1$. Representative data for system with linear size $L=96$ are shown in {Fig.~\ref{fig:2}{\bf a} and {\bf b}.}

With decreasing the temperature, the specific heat develops a broad hump, as shown in Fig.~\ref{fig:2}{\bf a}. In an AFM Heisenberg model ($Q=0$) on a square lattice, the peak of the specific heat characterizes a crossover below which short-range AFM correlations dominate. In the CBJQ model, we see that the hump becomes broader with increasing $Q$. This broadening indicates enhanced competition between the AFM and plaquette valence bond correlations. We have plotted the temperature corresponding to the specific heat peak in the phase diagram of Fig.~\ref{fig:1}{\bf b}. Away from $Q_c$, this crossover line increases almost linearly with $Q$. It shows a kinked feature when $Q$ is close to $Q_c$. This implies drastic changes of both ground states and low-energy excitations across the QPT, whose effects are detectable at finite temperature. {For $Q>Q_c$, the specific heat curve develops a weak peak at a temperature $T_{\rm p}$ below the broad hump (see inset of Fig.~\ref{fig:2}{\bf a}). Across this temperature, the system undergoes a second-order thermal transition to the PVBS. The transition temperature $T_{\rm p}$ can be accurately determined from the finite-size scaling of the Binder cumulant $U_p$ associated with the PVBS order parameter, as shown in Fig.~\ref{fig:2}{\bf c}.
}

The temperature dependent $1/T_1$ curve also exhibits rich crossover behavior. With decreasing $T$, $1/T_1$ increases, forming a broad peak at $T_{\rm{max}}$, similar to the behavior of the specific heat. Further decreasing $T$, $1/T_1$ decreases. For $Q<Q_c$ where the ground state is AFM, $1/T_1$ soon reaches a minimum at $T_{\rm{min}}$ and then increases rapidly, eventually divergent at $T\rightarrow0$. For $Q>Q_c$ where the ground state is a PVBS, $1/T_1$ decreases monotonically at low temperature and vanishes at $T\rightarrow0$. Interestingly, we observe that $1/T_1$ almost keeps a constant at low temperatures when $Q=Q_c\approx4.62$. Note that the $Q_c$ value from our $1/T_1$ data is slightly larger than that previously determined~\cite{Zhao_NP_2019}, which is most likely due to finite size effects. $T_{\rm{max}}$ and $T_{\rm{min}}$ define two more crossover lines as shown in Fig.~\ref{fig:1}{\bf b}. The high-temperature crossover line $T_{\rm{max}}$ follows the crossover defined by $C(T)$, below which the short-range AFM and/or plaquette valence bond correlations dominate. The low-temperature crossover $T_{\rm{min}}$, on the other hand, appears only on the AFM side. Below this crossover temperature, $1/T_1$ rises up and diverges in the $T\rightarrow0$ limit, implying that the system is approaching the long-range AFM order with a rapidly increasing correlation length. Therefore, $T_{\rm{min}}$ characterizes a crossover to the renormalized classical regime. As shown in the phase diagram of Fig.~\ref{fig:1}{\bf b}, the temperature range between $T_{\rm{min}}$ and $T_{\rm{max}}$ increases with increasing $Q$, especially when the system is close to the QPT. For a conventional continuous QPT, this regime is called quantum critical in which power-law scaling of thermodynamic and spectral quantities controlled by the QCP is observed. Such a scenario may not apply to a generic first-order transition. However, in the following we show that quantum scaling exists at the first-order QPT of the CBJQ model.

\section{Quantum scaling of the NMR spin-lattice relaxation rate}\label{Sec:Scaling}

\begin{figure*}[th!]
\centering\includegraphics[
width=1.0\textwidth
]{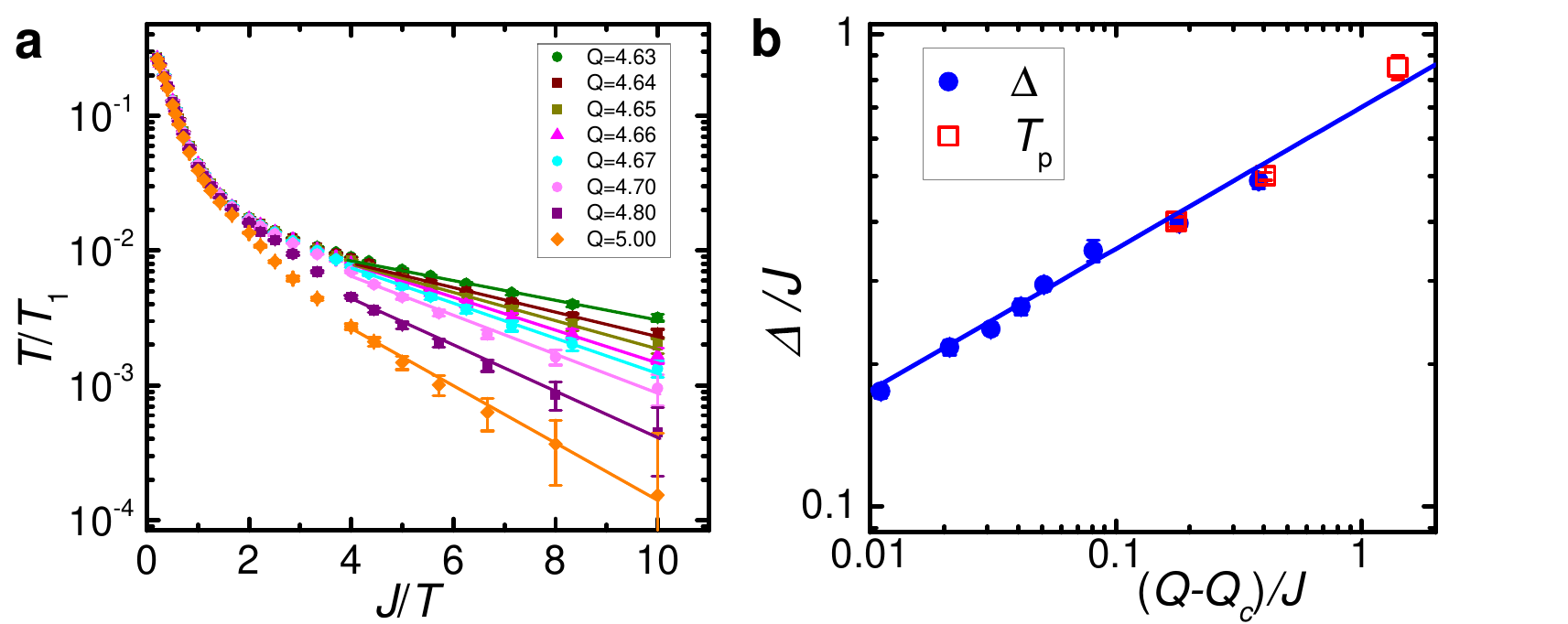}
\caption{(Color online) {\bf a} $1/T_1$ versus the inverse temperature for several $Q$ values. Lines are fits of $1/T_1\sim \frac{1}{T} e^{-\Delta/T}$ in the semi-log scale. {\bf b} Fitted spin gap $\Delta$ from panel {\bf a} (blue circles) and the transition temperature $T_{\rm p}$ to the PVBS phase (open squares) as a function of the distance to the QPT $Q-Q_c$ in the log-log scale. The line is a fit of $\Delta\sim (Q-Q_c)^{\phi}$ with the transition point $Q_c/J=4.619$ and the exponent $\phi=0.3$.
}
\label{fig:3}
\end{figure*}

We first examine the scaling property of $1/T_1$ for $Q>Q_c$. The ground state is a PVBS with a finite spin gap. As shown in Fig.~\ref{fig:3}{\bf a}, the low-temperature behavior follows $1/T_1\sim \frac{1}{T} e^{-\Delta/T}$, where $\Delta$ is the spin gap of the PVBS ground state. {As shown in Fig.~\ref{fig:2}{\bf d}, the finite-size effect effect of the $1/T_1$ curve is weak. We then study the spin gap of the largest system $L=96$.} The extracted gaps {at different $Q$ values} are plotted as blue circles in the phase diagram of Fig.~\ref{fig:1}{\bf b}. The gap reduces with decreasing $Q$ and can be fit by $\Delta\sim (Q-Q_c)^\phi$ near the transition point $Q_c$, as shown in Fig.~\ref{fig:3}{\bf b}. The best fit parameters are $Q_c=4.619\pm0.006$ and $\phi=0.30\pm0.05$. The fitted $Q_c$ value is consistent with the one determined from the flatness of the $1/T_1$ data, and $\phi\approx0.3$ implies that the gap closes at the transition in an unusual way because for a conventional first-order transition between two ordered states, the spin gap either closes abruptly or vanishes linearly with the applied magnetic field. Interestingly, for the transition in the CBJQ model, previous study found $\nu\approx 0.3$.\cite{Zhao_NP_2019} By assuming $z=1$ at this transition, we observe that $\phi\approx\nu z$, which usually holds for a continuous QPT. This unusual gap closing behavior is associated with the emergent enhanced symmetry at the transition, which reorganizes low-energy spin fluctuations by binding the AFM and PVBS order parameters to an $O(4)$ vector.

We then plot the transition temperature $T_{\rm{p}}$ in the same log-log plot of Fig.~\ref{fig:3}{\bf b}. Remarkably, the same scaling form of the spin gap $\Delta$ with $\phi\approx0.3$ also fits to $T_{\rm{p}}$ data. This confirms the above scenario that the low-energy spin fluctuations that disturb the PVBS order are indeed controlled by the emergent enhanced $O(4)$ symmetry at the transition point.

\begin{figure*}[th!]
\centering\includegraphics[
width=1.0\textwidth
]{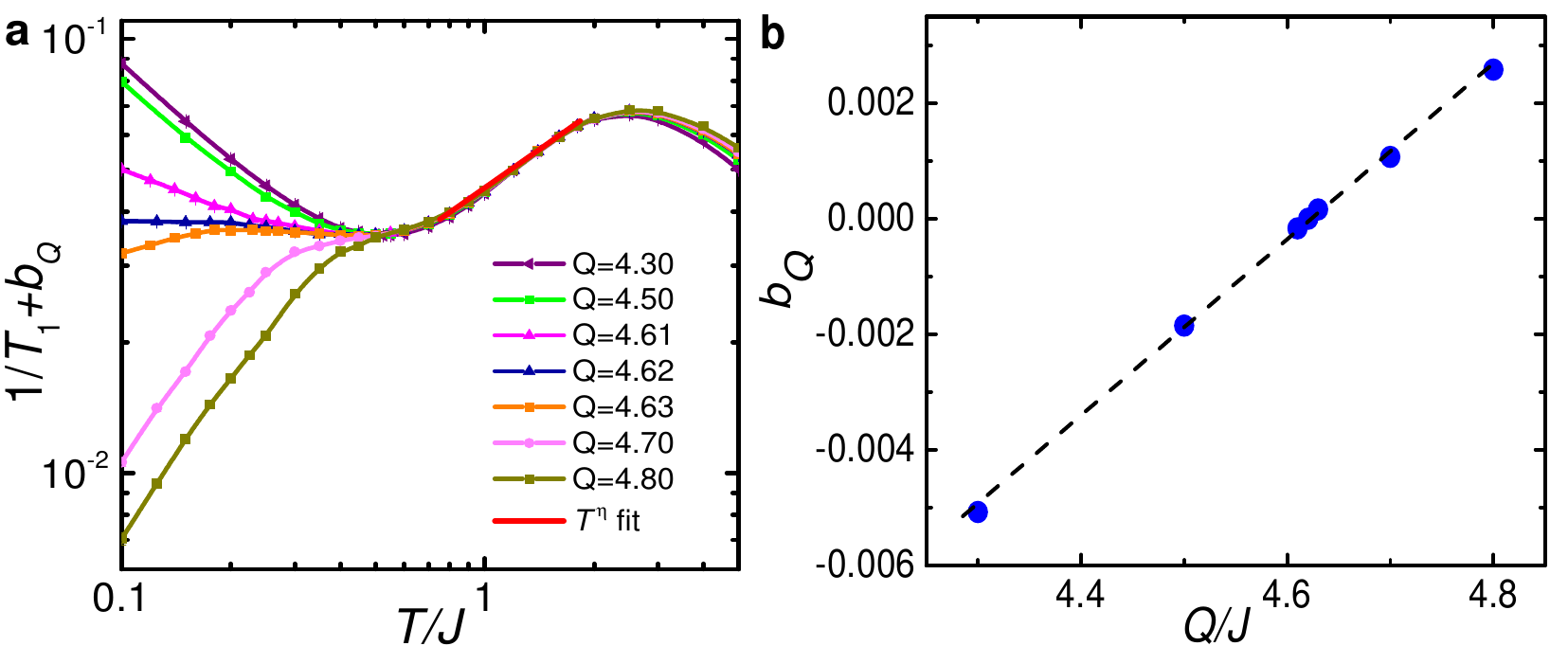}
\caption{(Color online) {\bf a} Scaling of $1/T_1$ with temperature. Data collapse between $T_{\rm{max}}$ and $T_{\rm{min}}$ implies a universal scaling of $1/T_1= a T^{\eta} - b_Q$ with $\eta\approx0.6$ (red line). {\bf b} $Q$ dependent of the non-universal constant $b_Q$. The dashed line is a linear fit.
}
\label{fig:4}
\end{figure*}

Despite its first-order nature, this transition shares many properties as a continuous QPT. We then wonder whether quantum-critical-like behaviors emerge. As we already discussed, the system within the temperature range $T_{\rm{min}}<T<T_{\rm{max}}$ behaves similar to the quantum critical regime. We then examine the scaling property of $1/T_1$ in this temperature region. Before we discuss the results near the transition point $Q_c$, we first take a look at the case $Q=0$, where the system is reduced to an AFM Heisenberg model. As shown in Fig.~\ref{fig:2}{\bf b}, $1/T_1$ is almost flat between the minimum and maximum for $0.3\lesssim T/J \lesssim 0.7$, and can be fit well by $1/T_1\sim T^\eta$ with $\eta\approx0.03$. This scaling behavior confirms a previous QMC result~\cite{Sandvik_PRB_1995} and the estimated $\eta$ is consistent with the value in the quantum critical regime of the $(2+1)D$ AFM Heisenberg model~\cite{Chubukov_PRB_1994}.

We now focus on the scaling of $1/T_1$ in the vicinity of the transition point $Q_c$ in the CBJQ model. As shown in Fig.~\ref{fig:4}{\bf a}, good data collapse is obtained by shifting the $1/T_1$ data with a temperature independent constant $b_Q$. In the temperature range between $T_{\rm{min}}$ and $T_{\rm{max}}$ (for $0.5\lesssim T/J \lesssim 2$), we find the following scaling form: $1/T_1= a T^\eta - b_Q$. Here $b_Q$ depends weakly on $Q$ (see Fig.~\ref{fig:4}{\bf b}) and $b_Q=0$ for $Q=Q_c$. $a$ is a non-universal coefficient and the $\eta$ is a universal exponent $\eta\approx0.6$. Similar to a continuous QPT, we see that $1/T_1\sim T^{\eta}$ at $Q=Q_c$ of the CBJQ model. But rather different from the quantum critical behavior for a continuous QPT, the universal scaling shows up above a finite temperature $T^\star\approx0.5$. Note that $T^\star$ is close to the crossover temperature $T_{\rm{min}}(Q\rightarrow Q_c)$. Below $T^\star$, $1/T_1$ exhibits distinct behaviors depending on $Q$. In particular for $Q=Q_c\approx4.62$, $1/T_1$ is very flat at low temperature. We have tried to fit to the low-temperature data at $Q=4.62$ with $T^\eta$, and get $\eta<0.001$. {Such a small effective $\eta$ value is inconsistent with that of a $(2+1)D$ $O(4)$ universality, for which $\eta\approx0.03$.\cite{Antonenko_PRE_1995} This implies that $1/T_1$ converges to a constant value at $T\rightarrow0$. Nonetheless, we cannot fully rule out a small but finite $\eta$ exponent due to the limitation of $T/J\geqslant0.1$ in our simulation.}

\section{Discussions and conclusion}\label{Sec:Conclusion}

Our results on the spin dynamics reveal the very unusual nature of the first-order QPT in the CBJQ model. Although the order parameters jump across this transition at $T=0$, the temperature dependent $1/T_1$ exhibits interesting quantum scaling similar to critical behaviors of a continuous QPT. This is seen from the gap closing $\Delta\sim (Q-Q_c)^\phi$ with the exponent $\phi\approx0.3\approx\nu z$, and from the scaling of $1/T_1 = a T^\eta - b_Q$ with the universal exponent $\eta\approx0.6$. One possible scenario accounting for this is that the emergent enhanced symmetry not only binds the AFM and PVBS order parameters to an $O(4)$ vector, but also reorganizes the low-energy spin fluctuations near the QPT that affect the spin dynamics. For example, an additional Goldstone mode associated with the enhanced $O(4)$ symmetry should emerge at the QPT. The appearance of this mode may explain the observed gap closing. Moreover, we expect fractionalized spin excitations form a continuum in the excitation spectrum with energies above the Goldstone mode, which may account for the quantum-critical-like behavior. It would be important to further investigate the full excitation spectrum of the model near the QPT. Nonetheless, here we stress that these effective exponents suggest a unique universality for the firs-order QPT with emergent $O(4)$ symmetry. As one can see obviously that $\nu\approx1/(d+z)$ and the unusually large $\eta$ are inconsistent with those of a continuous QPT with $O(N)$ symmetry. These effective exponents also differ from those at a DQCP with either $O(4)$ or $SO(5)$ emergent symmetry~\cite{Sandvik_PRL_2007, Qin_PRX_2017}. It is interesting to explore the connection between the quantum scaling behavior in the CBJQ model and that in models holding DQCPs.

We then address the implication for experimental results. In a recent NMR experiment~\cite{Cui_Science_2023}, quantum scaling of $1/T_1\sim T^\eta$ has been observed near a field induced proximate DQCP and the extracted effective exponent $\eta\approx0.2$. Such a large $\eta$ exponent cannot be understood by conventional $O(N)$ universality, but is consistent with that of DQCP with either $O(4)$ or $SO(5)$ emergent symmetry~\cite{Sandvik_PRL_2007, Qin_PRX_2017}. The experimental $\eta$ value is smaller than the value determined in our work. Here we point several factors that may be important for this difference. First, it is generally believed that the experimental material is well described by the Shastry-Sutherland model. Though the CBJQ model can also realize the AFM to PVBS transition, it is still an open question whether the strong spin frustration may modify these effective exponents. Second, the transition is induced by the magnetic field in the experiment. The applied field breaks the spin $SU(2)$ symmetry. Though it has been shown that an approximate $O(3)$ symmetry still holds at the field-induced AFM-PVBS QPT in the CBJQ model, the different symmetry may cause a change in the $\eta$ exponent. To get more insight, it would be interesting to study the spin dynamics of the CBJQ model under a magnetic field.

In conclusion, we have studied the low-energy spin dynamics of the CBJQ model by using quantum Monte Carlo simulations. The calculated NMR spin lattice relaxation rate $1/T_1$ exhibits rich finite-temperature crossover behavior. When the ground state is a PVBS, we can extract a finite spin gap $\Delta$ from the low-temperature $1/T_1$ data, which scales as $\Delta\sim (Q-Q_c)^\phi$ with the exponent $\phi\approx0.3$. We further find a universal scaling $1/T_1\sim T^{\eta}$ with $\eta\approx0.6$ near the AFM-PVBS QPT, suggesting a quantum scaling regime associated with the emergent enhanced symmetry at this first-order QPT.

We thank Y. Cui, A. Sandvik, N. Xi, and W. Yu for
useful discussions. This work is supported by the National Key R\&D Program of China (Grant No. 2023YFA1406500), and the National Science Foundation of China (Grant Nos. 12334008 and 12174441).


\section*{References}

\begin{thebibliography}{99}


\bibitem{Diep_book_2004} Diep H T 2004 Frustrated Spin Systems \href{https://books.google.com.sg/books?hl=zh-CN&lr=&id=Y4-6CgAAQBAJ&oi=fnd&pg=PR5&dq=frustrated+spin+systems&ots=XZCma5gDm6&sig=RMd-hqOi8Ckb_A5EslgMIrm-Imw&redir_esc=y#v=onepage&q=frustrated%20spin%20systems&f=false}{(World Scientific/Singapore)}

\bibitem{Mila_book_2010} Lacroix C, Mendels P, and Mila F  2010 Introduction to Frustrated Magnetism, Materials, Experiments, Theory \href{https://books.google.com.sg/books?hl=zh-CN&lr=&id=utSV09ZuhOkC&oi=fnd&pg=PR3&dq=Introduction+to+frustrated+magnetism,materials,experiments&ots=28bXLuVJUC&sig=NJMqVS3Re19HZvuKcHhR6KXxbXg&redir_esc=y#v=onepage&q=Introduction%20to%20frustrated%20magnetism%2Cmaterials%2Cexperiments&f=false}{(Springer/Berlin)}

\bibitem{Balents_Nature_2010} Balents  L 2010 Spin liquid in frustrated magnets \textit{Nature} \href{https://www.nature.com/articles/nature08917}{\textbf{464} 199}

\bibitem{Savary_RPP_2017} Savary L and Balents L 2017 Quantum spin liquids: A review \textit{Rep. Prog. Phys} \href{https://iopscience.iop.org/article/10.1088/0034-4885/80/1/016502/meta}{\textbf{80} 016502}

\bibitem{Zhou_RMP_2017}  Zhou Y,Kanoda K, and Ng T-K
2017 Quantum spin liquid states \textit{Rev. Mod. Phys} \href{https://journals.aps.org/rmp/abstract/10.1103/RevModPhys.89.025003}{\textbf{89} 025003}


\bibitem{Senthil_science_2004} Senthil T, Vishwanath A, Balents L, Sachdev S,
and Fisher M P A  2004 Deconfined Quantum Critical Points \textit{Science}
\href{https://www.science.org/doi/abs/10.1126/science.1091806}{\textbf{303} 1490}

\bibitem{Senthil_PRB_2004} Senthil T, Balents L, Sachdev S, Vishwanath A,
and Fisher M P A 2004 Quantum criticality beyond the Landau-Ginzburg-Wilson paradigm \textit{Phys. Rev. Lett}
\href{https://journals.aps.org/prb/abstract/10.1103/PhysRevB.70.144407}{ \textbf{70} 144407}

\bibitem{Nahum_PRX_2015} Nahum A, Chalker J T, Serna P, Ortu\~{n}o M, and Somoza A M 2015 Deconfined Quantum Criticality, Scaling Violations, and Classical Loop Models \textit{Phys. Rev. X}
\href{https://journals.aps.org/prx/abstract/10.1103/PhysRevX.5.041048}{ \textbf{5} 041048}


\bibitem{Nahum_PRL_2015} Nahum A, Serna P, Chalker J T, Ortu\~{n}o M, and Somoza A M 2015 Emergent SO(5) Symmetry at the Néel to Valence-Bond-Solid Transition \textit{Phys. Rev. Lett}
\href{https://journals.aps.org/prl/abstract/10.1103/PhysRevLett.115.267203}{ \textbf{115} 267203}


\bibitem{Shao_science_2016} Shao H, Guo W, and Sandvik A W 2016 Quantum criticality with two length scales \textit{Science}
\href{https://www.science.org/doi/full/10.1126/science.aad5007}{  \textbf{352} 213}

\bibitem{Wang_PRX_2017} Wang C, Nahum A, Metlitski M A, Xu C, and Senthil T 2017 Deconfined Quantum Critical Points: Symmetries and Dualities \textit{Phys. Rev. X}
\href{https://journals.aps.org/prx/abstract/10.1103/PhysRevX.7.031051}{ \textbf{7} 031051}


\bibitem{Ma_PRB_2018} Ma N, Sun G Y, You Y Z, Xu C, Vishwanath A,  Sandvik A W, and Meng Z Y 2018 Dynamical signature of fractionalization at a deconfined quantum critical point \textit{Phys. Rev. B}
\href{https://journals.aps.org/prb/abstract/10.1103/PhysRevB.98.174421}{ \textbf{98} 174421}

\bibitem{Lee_PRX_2019} Lee J Y, You Y Z, Sachdev S, and Vishwanath A 2019 Signatures of a Deconfined Phase Transition on the Shastry-Sutherland Lattice: Applications to Quantum Critical
$\mathrm{SrCu_2(BO_3)_2}$ \textit{Phys. Rev. X}
\href{https://journals.aps.org/prx/abstract/10.1103/PhysRevX.9.041037}{ \textbf{9}  041037}

\bibitem{Zhao_PRL_2020} Zhao B W, Takahashi J, and Sandvik A W 2020 Multicritical Deconfined Quantum Criticality and Lifshitz Point of a Helical Valence-Bond Phase \textit{Phys. Rev. Lett}
\href{https://journals.aps.org/prl/abstract/10.1103/PhysRevLett.125.257204}{  \textbf{125} 257204}


\bibitem{Sreejith_PRL_2019} Sreejith G J, Powell S, and Nahum A 2019 Emergent SO(5) Symmetry at the Columnar Ordering Transition in the Classical Cubic Dimer Model \textit{Phys. Rev. Lett}
\href{https://journals.aps.org/prl/abstract/10.1103/PhysRevLett.122.080601}{ \textbf{122} 080601}


\bibitem{Serna_PRB_2019} Serna P and Nahum A 2019 Emergence and spontaneous breaking of approximate
O(4) symmetry at a weakly first-order deconfined phase transition \textit{Phys. Rev. B}
\href{https://journals.aps.org/prb/abstract/10.1103/PhysRevB.99.195110}{ \textbf{99} 195110}

\bibitem{Takahashi_PRR_2020} Takahashi J and Sandvik A W 2020 Valence-bond solids, vestigial order, and emergent SO(5) symmetry in a two-dimensional quantum magnet \textit{Phys. Rev. Research}
\href{https://journals.aps.org/prresearch/abstract/10.1103/PhysRevResearch.2.033459}{  \textbf{2} 033459}


\bibitem{Zhang_PRB_2020} Zhang Y H and Sachdev S 2020 Deconfined criticality and ghost Fermi surfaces at the onset of antiferromagnetism in a metal \textit{Phys. Rev. B}
\href{https://journals.aps.org/prb/abstract/10.1103/PhysRevB.102.155124}{ \textbf{102}  155124}

\bibitem{Lu_PRB_2021} Lu D C, Xu C K, and You Y Z 2021 Self-duality protected multicriticality in deconfined quantum phase transitions \textit{Phys. Rev. B}
\href{https://journals.aps.org/prb/abstract/10.1103/PhysRevB.104.205142}{  \textbf{104} 205142}

\bibitem{Liu_SB_2022} Liu W Y, Gong S S, Li Y B, Poilblanc D, Chen W Q, and Gu Z C  2022 Gapless quantum spin liquid and global phase diagram of the spin-1/2 J1-J2 square antiferromagnetic Heisenberg model \textit{Sci. Bull.}
\href{https://www.sciengine.com/SB/doi/10.1016/j.scib.2022.03.010}{ \textbf{67}  1034}


\bibitem{Xi_PRB_2023} Xi N, Chen H Y, Xie Z Y, and Yu R 2023 Plaquette valence bond solid to antiferromagnet transition and deconfined quantum critical point of the Shastry-Sutherland model \textit{Phys. Rev. B}
\href{https://journals.aps.org/prb/abstract/10.1103/PhysRevB.107.L220408}{ \textbf{107} 220408}



\bibitem{Yang_PRB_2022} Yang J W, Sandvik A W, and Wang L 2022 Quantum criticality and spin liquid phase in the Shastry-Sutherland model \textit{Phys. Rev. B}
\href{https://journals.aps.org/prb/abstract/10.1103/PhysRevB.105.L060409}{ \textbf{105} L060409}

\bibitem{Liu_PRX_2022} Liu W Y, Hasik J, Gong S S, Poilblanc D, Chen W Q, and Gu Z C 2022 Emergence of Gapless Quantum Spin Liquid from Deconfined Quantum Critical Point \textit{Phys. Rev. X}
\href{https://journals.aps.org/prx/abstract/10.1103/PhysRevX.12.031039}{ \textbf{12} 031039}


\bibitem{Shackleton_PRB_2021} Shackleton H, Thomson A, and Sachdev S 2021 Deconfined criticality and a gapless $Z_2$ spin liquid in the square-lattice antiferromagnet \textit{Phys. Rev. B}
\href{https://journals.aps.org/prb/abstract/10.1103/PhysRevB.104.045110}{  \textbf{104} 045110}


\bibitem{Sandvik_PRL_2007} Sandvik A W 2007 Evidence for Deconfined Quantum Criticality in a Two-Dimensional Heisenberg Model with Four-Spin Interactions \textit{Phys. Rev. Lett}
\href{https://journals.aps.org/prl/abstract/10.1103/PhysRevLett.98.227202}{  \textbf{98} 227202}


\bibitem{Zhao_NP_2019} Zhao B W, Weinberg P, and Sandvik A W 2019 Symmetry-enhanced discontinuous phase transition in a two-dimensional quantum magnet \textit{Nat. Phys}
\href{https://www.nature.com/articles/s41567-019-0484-x}{ \textbf{15} 678}


\bibitem{Qin_PRX_2017} Qin Y Q, He Y Y, You Y Z, Lu Z Y, Sen A, Sandvik A W, Xu C K, and Meng Z Y 2017 Duality between the Deconfined Quantum-Critical Point and the Bosonic Topological Transition \textit{Phys. Rev. X}
\href{https://journals.aps.org/prx/abstract/10.1103/PhysRevX.7.031052}{ \textbf{7}  031052}

\bibitem{Sandvik_CPL_2020} Sandvik A W and Zhao B W 2020 Consistent Scaling Exponents at the Deconfined Quantum-Critical Point \textit{Chin. Phys. Lett}
\href{https://cpl.iphy.ac.cn/10.1088/0256-307X/37/5/057502#1}{ \textbf{37} 057502}

\bibitem{Cui_Science_2023} Cui, Y et al 2023 Proximate deconfined quantum critical point in  $\mathrm{SrCu_2(BO_3)_2}$ \textit{Science} \href{https://www.science.org/doi/abs/10.1126/science.adc9487}{\textbf{380} 1179}

\bibitem{Kageyama_PRL_1999} Kageyama H \textit{et al} 1999 Exact Dimer Ground State and Quantized Magnetization Plateaus in the Two-Dimensional Spin System $\mathrm{SrCu_2(BO_3)_2}$ \textit{Phys. Rev. Lett}
\href{https://journals.aps.org/prl/abstract/10.1103/PhysRevLett.82.3168}{ \textbf{82} 3168}

\bibitem{Miyahara_PRL_1999} Miyahara S and Ueda K  1999 Exact Dimer Ground State of the Two Dimensional Heisenberg Spin System $\mathrm{SrCu_2(BO_3)_2}$ \textit{Phys. Rev. Lett}
\href{https://journals.aps.org/prl/abstract/10.1103/PhysRevLett.82.3701}{ \textbf{82} 3701}

\bibitem{Miyahara_JPCM_2003} Miyahara S and Ueda K 2003 Theory of the orthogonal dimer Heisenberg spin model for $\mathrm{SrCu_2(BO_3)_2}$ \textit{J. Phys.: Condens. Matter}
\href{https://iopscience.iop.org/article/10.1088/0953-8984/15/9/201/meta}{  \textbf{15} R327}

\bibitem{Shastry_PBC_1981} Shastry B S and Sutherland B 1981 \textit{Physica B+C}
\href{http://physics.ucsc.edu/~sriram/papers/paper_19.pdf}{ \textbf{108} 1069}

\bibitem{Koga_PRL_2000} Koga A and Kawakami N 2000 Exact ground state of a quantum mechanical antiferromagnet \textit{Phys. Rev. Lett}
\href{https://journals.aps.org/prl/abstract/10.1103/PhysRevLett.84.4461}{ \textbf{84} 4461}

\bibitem{Corboz_PRB_2013} Corboz P and Mila F 2013 Tensor network study of the Shastry-Sutherland model in zero magnetic field \textit{Phys. Rev. B}
\href{https://journals.aps.org/prb/abstract/10.1103/PhysRevB.87.115144}{ \textbf{87} 115144}


\bibitem{Kodama_JPCM_2005} Kodama K, Miyahara S, Takigawa M, Horvati\'{c} M, Berthier C, Mila F , Kageyama H and Ueda Y  2005 Field-induced effects of anisotropic magnetic interactions in $\mathrm{SrCu_2(BO_3)_2}$ \textit{J. Phys.:Condens. Matter}
\href{https://iopscience.iop.org/article/10.1088/0953-8984/17/4/L02/meta}{ \textbf{17} L61}


\bibitem{Kageyama_JPSJ_1998} Kageyama H, Onizuka K, Ueda Y, Mushnikov N V, Goto T, Yoshimura K, and Kosuge K 1998 Magnetic Anisotropy of $\mathrm{SrCu_2(BO_3)_2}$ with a Two-Dimensional Orthogonal Dimer Lattice \textit{J. Phys. Soc. Jpn}
\href{https://www.jstage.jst.go.jp/article/jpsj/67/12/67_12_4304/_article/-char/ja/}{  \textbf{67} 4304}


\bibitem{Zayed_NP_2017} Zayed M E \textit{et al} 2017 4-spin plaquette singlet state in the Shastry–Sutherland compound $\mathrm{SrCu_2(BO_3)_2}$ \textit{Nat. Phys}
\href{https://www.nature.com/articles/nphys4190}{ \textbf{13} 962}

\bibitem{Guo_PRL_2020} Guo J \textit{et al} 2020 Quantum Phases of $\mathrm{SrCu_2(BO_3)_2}$ from High-Pressure Thermodynamics \textit{Phys. Rev. Lett}
\href{https://journals.aps.org/prl/abstract/10.1103/PhysRevLett.124.206602}{ \textbf{124} 206602}

\bibitem{Jimenez_Nature_2021} Jim\'{e}nez J L \textit{et al} 2021 A quantum magnetic analogue to the critical point of water \textit{Nature}
\href{https://www.nature.com/articles/s41586-021-03411-8}{  \textbf{592} 370}

\bibitem{Haravifard_NC_2016} Haravifard S, Graf D, Feiguin A E, Batista C D, Lang J C, Silevitch D M, Srajer G, Gaulin B D, Dabkowska  H A, and Rosenbaum T F 2016 \textit{Nat. Commun}
\href{https://www.nature.com/articles/ncomms11956}{ \textbf{7} 11956}



\bibitem{Shi_NC_2022} Shi Z Z, Dissanayake S, Corboz P, Steinhardt W, Graf D, Silevitch D M, Dabkowska H A, Rosenbaum T F, Mila Fr\'{e}d\'{e}ric, and Haravifard S 2022 \textit{Nat. Commun}  \href{https://www.nature.com/articles/s41467-022-30036-w}{ \textbf{13} 2301}


\bibitem{Keles_PRB_2022} Keleş A and Zhao E 2022 Rise and fall of plaquette order in the Shastry-Sutherland magnet revealed by pseudofermion functional renormalization group \textit{Phys. Rev. B}
\href{https://journals.aps.org/prb/abstract/10.1103/PhysRevB.105.L041115}{ \textbf{105} L041115}


\bibitem{Boos_PRB_2019} Boos C, Crone S P G, Niesen I A, Corboz P, Schmidt K P, and Mila F  2019 Competition between intermediate plaquette phases in $\mathrm{SrCu_2(BO_3)_2}$ under pressure \textit{Phys. Rev. B}
\href{https://journals.aps.org/prb/abstract/10.1103/PhysRevB.100.140413}{\textbf{100}  140413}


\bibitem{Wessel_PRB_2018} Wessel S, Niesen I, Stapmanns J, Normand B, Mila F,
Corboz P, and Honecker A 2018 Thermodynamic properties of the Shastry-Sutherland model from quantum Monte Carlo simulations \textit{Phys. Rev. B}
\href{https://journals.aps.org/prb/abstract/10.1103/PhysRevB.98.174432}{ \textbf{98} 174432}


\bibitem{Sun_CPB_2021} Sun G Y, Ma N S, Zhao B W, Sandvik A W, and Meng Z Y 2021 Emergent O(4) symmetry at the phase transition from plaquette-singlet to antiferromagnetic order in quasi-two-dimensional quantum magnets \textit{Chin. Phys. B}
\href{https://cpb.iphy.ac.cn/EN/10.1088/1674-1056/abf3b8}{ \textbf{30} 067505}

\bibitem{Dorneich_PRE_2001} Dorneich A and Troyer M  2001 Accessing the dynamics of large many-particle systems using the stochastic series expansion \textit{Phys. Rev. E}
\href{https://journals.aps.org/pre/abstract/10.1103/PhysRevE.64.066701}{\textbf{64} 066701}


\bibitem{Sandvik_AIPCP_2010} Sandvik A W 2010 Computational Studies of Quantum Spin Systems  \textit{AIP Conf. Proc}
\href{https://pubs.aip.org/aip/acp/article/1297/1/135/854814/Computational-Studies-of-Quantum-Spin-Systems}{ \textbf{1297} 135}

\bibitem{Syljuasen_PRE_2002} Sylju\.{a}sen O F and Sandvik A W  2002 Quantum Monte Carlo with directed loops \textit{Phys. Rev. E}
\href{https://journals.aps.org/pre/abstract/10.1103/PhysRevE.66.046701}{\textbf{66} 046701}

\bibitem{Randeria_PRL_1992} Randeria M, Trivedi N,  Moreo A, and Scalettar R T 1992 Pairing and spin gap in the normal state of short coherence length superconductors \textit{Phys. Rev. Let} \href{https://journals.aps.org/prl/abstract/10.1103/PhysRevLett.69.2001}{\textbf{69} 2001}


\bibitem{Fan_PRR_2020} Fan Y C, Yang J H, Yu W Q, Wu J D, and Yu R 2020 Phase diagram and quantum criticality of Heisenberg spin chains with Ising anisotropic interchain couplings \textit{Phys. Rev. Research} \href{https://journals.aps.org/prresearch/abstract/10.1103/PhysRevResearch.2.013345}{\textbf{2} 013345 }

\bibitem{Sandvik_PRB_1995} Sandvik A W and Scalapino D J 1995 Spin dynamics of $\mathrm{La_2CuO_4}$ and the two-dimensional Heisenberg model \textit{Phys. Rev. B}  \href{https://journals.aps.org/prb/abstract/10.1103/PhysRevB.51.9403} {\textbf{51} 9403}


\bibitem{Chubukov_PRB_1994} Chubukov A V, Sachdev S, and Ye J  1994 Theory of two-dimensional quantum Heisenberg antiferromagnets with a nearly critical ground state \textit{Phys. Rev. B}
\href{https://journals.aps.org/prb/abstract/10.1103/PhysRevB.49.11919}{ \textbf{49} 11919}

\bibitem{Antonenko_PRE_1995} Antonenko S A and Sokolov A I 1995 Critical exponents for a three-dimensional $O(n)$-symmetric model with $n>3$ \textit{Phys. Rev. E}
\href{https://journals.aps.org/pre/abstract/10.1103/PhysRevE.51.1894}{\textbf{51} 1894}

\end{thebibliography}
\end{document}